\documentclass[a4paper,12pt]{article}

\usepackage[centertags]{amsmath}
\usepackage{amsfonts}
\usepackage{amssymb}
\usepackage{amsthm}
\usepackage{newlfont}
\usepackage[english]{babel}
\usepackage{enumerate}

\newtheorem{theorem}{Theorem}

\theoremstyle{remark}

\theoremstyle{definition}

\theoremstyle{definition}

     \newcommand {\beq}  {\begin{equation}}
      \newcommand {\eeq}  {\end{equation}}

\author{J.P.Cruz$^*$\hspace{1cm}Lakshtanov E.L.\thanks{Department of
Mathematics, Aveiro University, Aveiro 3810, Portugal.  This work
was supported by {\it Centre for Research on Optimization and
Control} (CEOC) from the ``{\it Funda\c{c}\~{a}o para a
Ci\^{e}ncia e a Tecnologia}'' (FCT), cofinanced by the European
Community Fund FEDER/POCTI, and by the FCT research project
PTDC/MAT/72840/2006.} \thanks{e-mail: lakshtanov@rambler.ru}}

\title{Upper bounds for Total Cross Section  in scattering by an obstacle with impedance boundary conditions}

\begin{document}
\date{}
\maketitle


\begin{abstract}
The scalar scattering of a plane wave by a smooth obstacle with
impedance boundary conditions is considered.  Upper bounds for
the Total Cross Section and for the absorbed power are presented.
\end{abstract}

\section{Introduction}

Consider a bounded body $\Omega \subset \mathbb R^3$ with
Lipschitz boundary $\partial \Omega$ and $k > 0$. The scattered
field is given by the Helmholtz equation and a radiation condition
\begin{equation}\label{helm}
\Delta u(r)+k^2 u(r)=0, \quad r \in \Omega'=\mathbb R^3 \backslash
\Omega
\end{equation}
\begin{equation}\label{Somm}
\int_{|r|=R} \left |\frac{\partial u(r)}{\partial |r|}-iku(r)
\right |^2 dS = o(1), \quad R \rightarrow \infty,
\end{equation}
with impedance boundary conditions of the form
\begin{equation}\label{dir}
\left (\frac{\partial}{\partial n} + (a+ib) \right )
(u(r)+e^{ikz}) = 0,  \quad \quad r=(x,y,z) \in
\partial \Omega, \quad a,b \in \mathbb R, b>0
\end{equation}
 $e^{ikz}=e^{ik(r \cdot \theta_0)}$ is an incident field formed by
a plane wave with incident angle $\theta_0=(0,0,1) \in S^2$. In \cite{vantychonovsamarsky}, for example, is proved the existence and
uniqueness of the solution of (\ref{helm})-(\ref{dir}). A function
$u(r)$ which satisfies the mentioned conditions has asymptotic
\begin{equation}\label{scamp}
u(r)=\frac{e^{ik|r|}}{|r|} f_{a,b}(\theta)+o \left (\frac{1}{|r|}
\right ), \quad r \rightarrow \infty, \quad \theta=r/|r| \in S^2,
\end{equation}
where the function $f_{a,b}(\theta)=f_{a,b}(\theta,k)$ is called
{\it scattering amplitude} and the quantity
$$
\sigma_{a,b}=\int_{S^2} |f_{a,b}(\theta)|^2 d\sigma(\theta)
$$
is called the Total Cross Section. $\sigma$ is a square element of
the unit sphere.

The power absorbed by the obstacle is given by
$$
\mathcal L =b \|e^{ikz}+u\|^2_{L_2(\partial \Omega,dS)}
$$
It is valid the optical theorem (for example see \cite{Kriegsmann})
$$
 \sigma_{a,b} = \frac{4\pi}{k} Im [f_{a,b}(\theta_0)] + \frac{1}{k}\mathcal L.
$$

\begin{theorem}\label{maj}
The following inequalities hold for total cross section
\begin{equation}\label{majest}
 \sigma_{a,b} \leq \frac{S(k+\sqrt{a^2+b^2})^2(b+\sqrt{a^2+b^2})}{kb^2}
\end{equation}
and for the power absorbed by the obstacle
\begin{equation}\label{majest2}
\mathcal L \leq \frac{S(k+b+\sqrt{a^2+b^2})^2}{b}.
\end{equation}
\end{theorem}
Note that similar inequalities for Dirichlet and Neumann cases
($\gamma=\infty$) or ($\gamma=0$) are known only in the limit
cases $k=0$ or $k=\infty$ (see \cite{wl}).

In particulary these inequalities solve the question if for
certain $a+ib, b>0$ and $k$ there exists a sequence of smooth obstacles
with uniformly bounded area such that $\sigma_{a,b}$ tends to
infinity.
\subsection{Proofs}
Let $u$ be the field of the outgoing wave satisfying
(\ref{helm}),(\ref{Somm}),(\ref{dir}). Everywere $\| \cdot
\| =\| \cdot \|_{L_2(\partial \Omega,dS)}.$ Lets prove that
\begin{equation}\label{f1}
\left  \| \frac{\partial u}{\partial n}+(a+ib)u \right \| \geq b
\| u \|.
\end{equation}
Note that
$$
\left \| \frac{\partial u}{\partial n}+(a+ib)u \right \|^2 = \left
\| \frac{\partial u}{\partial n}+a u \right \|^2+2b Im \left (
\int_{\partial \Omega} \frac{\partial u}{\partial n} \overline{u}
dS \right )+b^2 \|u\|^2.
$$
Using the well known fact (evidence from the Second Green's
identity)
\begin{equation}\label{f2}
Im \left ( \int_{\partial \Omega} \frac{\partial u}{\partial n}
\overline{u} dS \right )=k\sigma_{a,b} \geq 0,
\end{equation}
we obtain (\ref{f1}). Note now that from (\ref{dir}) follows that
$$
\left \|\frac{\partial u}{\partial n}+(a+ib)u \right \| = \left
\|\frac{\partial e^{ikz}}{\partial n}+(a+ib)e^{ikz} \right \| \leq
$$
$$
\left \| \frac{\partial e^{ikz}}{\partial n} \right \| +
\sqrt{a^2+b^2}\|e^{ikz}\| \leq \sqrt{S}(k+\sqrt{a^2+b^2})
$$
So using (\ref{f1}) we obtain
\begin{equation}\label{f3}
b\|u\| \leq \sqrt{S} (k+\sqrt{a^2+b^2}),
\end{equation}
and, in particularly, (\ref{majest2}). Also from (\ref{dir}) we have for $r \in \partial \Omega$
$$
-\frac{\partial u}{\partial n}=(a+ib)u + \frac{\partial
e^{ikz}}{\partial n}+(a+ib)e^{ikz},
$$
therefore
$$
\left \|\frac{\partial u}{\partial n} \right \| \leq
\sqrt{a^2+b^2} \|u|+ \left \| \frac{\partial e^{ikz}}{\partial n}
\right \| + \sqrt{a^2+b^2}\|e^{ikz}\| \leq
$$
$$
\sqrt{a^2+b^2}\|u\|+\sqrt{S}(k+\sqrt{a^2+b^2}) \leq
\frac{\sqrt{S}(k+\sqrt{a^2+b^2})(b+\sqrt{a^2+b^2})}{b}.
$$
Now from (\ref{f2}) we have
$$
\sigma_{a,b}\leq \frac{1}{k} \|u\| \|\frac{\partial u}{\partial
n}\| \leq
 \frac{1}{k} \left (\frac{\sqrt{S} (k+\sqrt{a^2+b^2})}{b}
\right )\cdot
$$
$$
\left (\frac{\sqrt{S}(k+\sqrt{a^2+b^2})(b+\sqrt{a^2+b^2})}{b}
\right )=
 \frac{{S}(k+\sqrt{a^2+b^2})^2(b+\sqrt{a^2+b^2})}{kb^2}
$$

 Theorem \ref{maj} is proved.

\end{document}